\documentclass[reqno,centertags, 12pt]{amsart}
\usepackage{amsmath,amsthm,amscd,amssymb}
\usepackage{latexsym}
\usepackage{curves}    

\newcommand{\bbR}{{\mathbb{R}}}

\newcommand{\bbC}{{\mathbb{C}}}

\newcommand{\calE}{{\mathcal E}}

\newcommand{\lb}{\label}
\newcommand{\f}{\frac}

\newcommand{\ess}{\text{\rm{ess}}}

\newcommand{\bi}{\bibitem}

\newcommand{\beq}{\begin{equation}}
\newcommand{\eeq}{\end{equation}}
\newcommand{\ba}{\begin{align}}
\newcommand{\ea}{\end{align}}
\newcommand{\eps}{\varepsilon}
\newcommand{\del}{\delta}
\newcommand{\tht}{\theta}
\newcommand{\al}{\alpha}
\newcommand{\be}{\beta}
\newcommand{\partt}{\tfrac{\partial}{\partial t}}
\newcommand{\lan}{\langle}
\newcommand{\ran}{\rangle}
\newcommand{\til}{\tilde}

\newcounter{smalllist}
\newenvironment{SL}{\begin{list}{{\rm\roman{smalllist})}}{%
\setlength{\topsep}{0mm}\setlength{\parsep}{0mm}\setlength{\itemsep}{0mm}%
\setlength{\labelwidth}{2em}\setlength{\leftmargin}{2em}\usecounter{smalllist}%
}}{\end{list}}

\DeclareMathOperator{\sgn}{sgn}

\allowdisplaybreaks
\numberwithin{equation}{section}

\newtheorem{theorem}{Theorem}[section]

\newtheorem{proposition}[theorem]{Proposition}
\newtheorem{lemma}[theorem]{Lemma}
\newtheorem{corollary}[theorem]{Corollary}
\theoremstyle{definition}
\newtheorem{definition}[theorem]{Definition}

\theoremstyle{remark}

\newcommand{\abs}[1]{\lvert#1\rvert}

\begin{document}
\title[Szeg\H{o} Condition for Coulomb Jacobi Matrices]{\centerline{The Szeg\H{o} Condition for} \centerline{Coulomb Jacobi Matrices}}
\author[Andrej Zlato\v{s}]{Andrej Zlato\v{s}}

\address{Mathematics 253-37 \\
California Institute of Technology \\
Pasadena, CA 91125, USA }

\email{andrej@caltech.edu}

\thanks{2000 {\it Mathematics Subject Classification}. Primary: 42C05; Secondary: 47B36}

\thanks{{\it Key words}. Jacobi matrix, Szeg\H o condition, Coulomb perturbations}

\date{July 24, 2002}

\begin{abstract} 
A Jacobi matrix with $a_n\to 1$, $b_n\to 0$ and spectral measure $\nu'(x)dx+d\nu_{sing}(x)$ satisfies the Szeg\H o condition if
\[
\int_{0}^\pi \ln \bigl[ \nu'(2\cos\tht)\bigr]d\tht   
\text{\phantom{iiiiiiiiiiiiiiiiiiiiiiii}}
\]
is finite. We prove that if
\[
a_n\equiv 1+\frac \al n + O(n^{-1-\eps}) \qquad \qquad b_n\equiv \frac \be n +O(n^{-1-\eps})
\text{\phantom{iiiiiiiiiiiiiiiiiiiiiiiii}}
\]
with $2\al\ge |\be|$ and $\eps>0$, then the corresponding matrix is Szeg\H o.
\end{abstract}

\maketitle

\section{Introduction} \lb{S1}

In this paper we discuss the Szeg\H o condition for Jacobi matrices and orthogonal polynomials. A Jacobi matrix is the matrix
\begin{equation} \lb{1.0}
J= \begin{pmatrix} 
b_1 & a_1 & 0 & \dots \\
a_1 & b_2 & a_2 & \dots \\
0 & a_2 & b_3 & \dots \\
\dots & \dots & \dots & \dots \\
\end{pmatrix}
\end{equation}
with $a_n>0$ and $b_n\in\bbR$. We let $\nu$ be the spectral measure of $J$ as an operator on $\ell^2(\{ 0,1,\dots \})$, with respect to the vector $\delta_0$. That is,
\[
\langle\delta_0,(J-z)^{-1}\delta_0\rangle=\int\frac{d\nu(x)}{x-z}
\]
for $z\in\bbC\smallsetminus\bbR$.
We denote $\{P_n(x)\}_{n=0}^\infty$ the orthonormal polynomials for this measure, obtained from $\{x^n\}_{n=0}^\infty$ by the Gram-Schmidt procedure. Since $\nu(\bbR)=\|\delta_0\|^2=1$, we have $P_{0}(x)\equiv 1$. If we define $P_{-1}(x)\equiv 0$, then the $P_n$'s obey the three term recurrence relation for $n\ge 0$
\begin{equation} \lb{1.0a}
xP_n(x)=a_{n+1}P_{n+1}(x)+b_{n+1}P_n(x)+a_nP_{n-1}(x) 
\end{equation}
Hence, $\{P_n(x)\}_{n=0}^\infty$ is the Dirichlet eigenfunction of $J$ for energy $x$. This relationship establishes a one-to-one correspondence between bounded Jacobi matrices and polynomials orthonormal w.r.t.~measures with bounded infinite support and total mass 1.

We will usually consider $J$ such that $J-J_0$ is compact. Here $J_0$ is the free Jacobi matrix with $a_n\equiv 1$, $b_n\equiv 0$ and $d\nu_0(x)=(2\pi)^{-1}\chi_{[-2,2]}\sqrt{4-x^2}dx$. For such $J$ we have $a_n\to 1$ and $b_n\to 0$ and $\sigma_{\ess}(J)=[-2,2]$. Outside of this interval $J$ can only have simple isolated eigenvalues, with $\pm 2$ the only possible accumulation points. We denote them $E_1^+>E_2^+>\cdots> 2$ and $E_1^-<E_2^-<\cdots< -2$.

The main object of our interest is the {\it Szeg\H o integral}
\begin{equation} \lb{1.1}
Z(J)\equiv \f{1}{2\pi} \int_{-2}^2 \ln \biggl( \f{\sqrt{4-x^2\,}}{2\pi \nu'(x)}\biggr)
\f{dx}{\sqrt{4-x^2}}
\end{equation}
where $\nu'(x)\equiv d\nu_{ac}(x)/dx$. We say that $J$ satisfies the {\it Szeg\H o condition\/} if $Z(J)$ is finite. It can be proved that the negative part of the integral in \eqref{1.1} is always integrable and $Z(J)\ge-\frac 12\ln(2)$ (see Killip-Simon \cite{KS}). Hence, we are left with the question whether $Z(J)<\infty$. There is extensive literature on when this is the case (e.g.~\cite{AI,Cas,DN,Gon,Nev1,Nev2,Nik,PY,Sho,SZ,Sze}), and so one is interested in answering this question.

Notice that 
\[
-2\pi Z(J)= \int_{0}^\pi \ln \biggl( \f{\pi\nu'(2\cos\tht)}{\sin\tht}\biggr)d\tht = \int_{0}^\pi \ln \bigl( \nu'(2\cos\tht)\bigr)d\tht + const.
\]
with $x=2\cos\tht$. Many authors consider the last integral instead of $Z(J)$ and the question is whether $\int_{0}^\pi \ln \bigl( \nu'(2\cos\tht)\bigr)d\tht>-\infty$. For our purposes, $Z(J)$ is more suitable. Also notice that $Z(J)<\infty$ implies that the essential support of $\nu_{ac}$ is $[-2,2]$.

In this paper, we want to address a conjecture of Askey about Coulomb-type Jacobi matrices, reported by Nevai in \cite{Nev1}. Askey conjectured that if
\begin{equation}  \lb{1.2}
a_n\equiv 1+\frac \al n +O\left(\frac 1{n^2}\right) \qquad \qquad b_n\equiv \frac \be n +O\left(\frac 1{n^2}\right)
\end{equation}
with $(\al,\be)\neq (0,0)$, then the Szeg\H o condition fails (it has been known that it holds if $\alpha=\beta=0$).
Later, however, Askey-Ismail \cite{AI} found some explicit examples with $b_n\equiv 0$ and $\al>0$ for which the Szeg\H o condition holds! And in \cite{DN}, Dombrowski-Nevai proved that the condition holds whenever $b_n\equiv 0$ and $a_n\equiv 1+\al/n+o(n^{-2})$ with $\al>0$. In conclusion, the conjecture had to be modified.

The ``right'' form of the conjecture can be guessed from Charris-Ismail \cite{CI}, who computed the weights for certain Pollaczek-type polynomials (with $a_n$, $b_n$ of the form \eqref{1.2}). Although they did not note it, their examples are Szeg\H o if and only if $2\al\ge|\be|$. We will see that this is true in general.

The first result which allows errors of the type \eqref{1.2} was proved by Simon and the author in \cite{SZ}, and is in-line with this picture. Indeed, the following appears in \cite{SZ}.

\begin{proposition} \lb{P1.1} If 
\begin{equation}  \lb{1.3}
a_n\equiv 1+\frac \al n +E_a (n) \qquad \qquad b_n\equiv \frac \be n +E_b (n)
\end{equation}
with
\[
\sum_{n=1}^N \bigr(\abs{E_a(n)} + \abs{E_b (n)}\bigl) =o\bigl(\ln(N)\bigr)
\]
and $2\al<|\be|$, then the Szeg\H{o} condition fails. 
\end{proposition}

So Askey was right in the case $2\al<|\be|$. The present paper concentrates on the complementary region $2\al\ge |\be|$ and shows that the Szeg\H o condition holds there (see figure below). Here is our main result. We denote $a_+\equiv \max\{a,0\}$ and $a_-\equiv -\min\{a,0\}$.

\begin{theorem}[=Theorem \ref{T4.3}] \lb{T1.2}
Let
\begin{equation}  \lb{1.4}
a_n\equiv c_n + O(n^{-1-\eps}) \qquad \qquad b_n\equiv d_n +O(n^{-1-\eps})
\end{equation}
for some $\eps>0$, where $c_n\ge 1+\frac{|d_n|}2$ for $n>N$, $\lim_{n\to\infty} c_n=1$ and
\begin{equation}  \lb{1.5}
\sum_{n=1}^\infty n\biggl[c_{n+1}^2-c_{n}^2 + \frac{c_{n+1}}2|d_{n+2}-d_{n+1}| + \frac{c_{n}}2|d_{n+1}-d_{n}|\biggr]_+<\infty
\end{equation}
Then the matrix $J$, given by \eqref{1.0}, satisfies the Szeg\H o condition.
\end{theorem}

{\it Remarks.}
1. Notice that the sum in \eqref{1.5} cannot be simplified. We cannot replace the last two terms by $c_n|d_{n+1}-d_n|$ because we take positive parts of the summands in \eqref{1.5}.
\smallskip

2. In particular one can take $c_n\equiv 1+\al/n$ and $d_n\equiv \be/n$ with $2\al\ge |\be|$.
\smallskip

We will prove this theorem in two steps. The first one is an extension of the result in \cite{DN} and shows that $J$ is Szeg\H o whenever $a_n, b_n$ satisfy the conditions for $c_n, d_n$ in Theorem \ref{T1.2}. 

The second step lets us add $O(n^{-1-\eps})$ errors to such $c_n,d_n$. Our tool here are the Case sum rules for Jacobi matrices, in particular the {\it step-by-step $Z$ sum rule\/} \eqref{1.6} below (called $C_0$ in \cite{KS}). These were introduced by Case \cite{Cas}, recently extended in \cite{KS}, and finally proved in the form we use here in \cite{SZ} (see \cite{DK,Den} for related Schr\" odinger operators results).  
We let $\beta_j^\pm$ be such that $E_j^\pm=\beta_j^\pm+(\beta_j^\pm)^{-1}$ and $\pm\beta_j^\pm>1$. If $J$ has fewer than $j$  eigenvalues above $2$/below $-2$, we define $\beta_j^{+/-}\equiv +1/-1$. Let $J^{(n)}$ be the matrix obtained from $J$ by removing $n$ top rows and $n$ leftmost columns. It was proved in \cite{SZ} that if $J-J_0$ is compact, then we have 
\begin{equation} \lb{1.6}
Z(J) = -\sum_{j=1}^n \ln (a_j) + \sum_{\pm}\sum_{j}\Big(\ln|\be_j^\pm(J)|-\ln|\be_j^\pm(J^{(n)})|\Big) + Z(J^{(n)}) 
\end{equation}
and that the double sum is always convergent with non-negative terms. 

\eqref{1.6} holds even if $Z(J)=\infty$, and so $J$ is Szeg\H o if and only if $J^{(n)}$ is. In particular, the Szeg\H o condition is stable under finite-rank perturbations. We will be able to pass to certain infinite-rank perturbations of $J$ by representing them as limits of finite-rank perturbations and using lower semicontinuity of $Z$ in $J$ proved in \cite{KS}. To do this, we will need to control the change of the $E_j^\pm$'s under these perturbations, in order to estimate the double sum in \eqref{1.6} (or, more precisely, in \eqref{4.0a} below).

The rest of the paper is organized as follows. In Section \ref{S2} we extend the above mentioned result from \cite{DN}. In Section \ref{S3} we prove results on the control of change of eigenvalues under certain finite-rank perturbations. In Section \ref{S4} we use these to prove Theorem \ref{T1.2}, along with some related results. 

Finally, Section \ref{S5} discusses some situations when the Szeg\H o integral is allowed to diverge at one end ({\it one-sided Szeg\H o conditions}). We study the case \eqref{1.2} with $O(n^{-1-\eps})$ errors and establish the following picture. 

\vskip 5mm

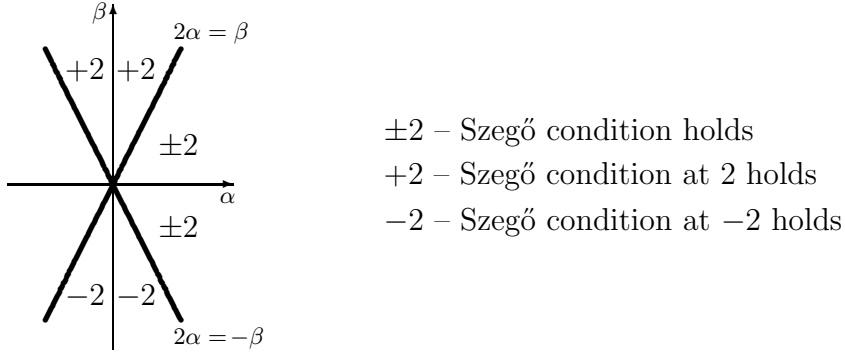
\setlength{\unitlength}{0.2mm}
\begin{picture}(450,230)

\thinlines
\put(0,110){\vector(1,0){150}}
\put(70,0){\vector(0,1){230}}
\linethickness{2pt}
\curve(25,20,115,200)
\curve(25,200,115,20)

\put(141,98){\text{\scriptsize $\alpha$}}
\put(56,220){\text{\scriptsize $\beta$}}
\put(110,207){\text{\scriptsize $2\alpha=\beta$}}
\put(110,4){\text{\scriptsize $2\alpha=\! -\beta$}}
\put(100,75){\text{$\pm 2$}}
\put(100,130){\text{$\pm 2$}}
\put(72,30){\text{$-2$}}
\put(72,180){\text{$+2$}}
\put(38,30){\text{$-2$}}
\put(38,180){\text{$+2$}}

\put(250,140){\text{{$\pm 2$} -- Szeg\H o condition holds}}
\put(250,110){\text{{$+2$} -- Szeg\H o condition at $2$ holds}}
\put(250,80){\text{{$-2$} -- Szeg\H o condition at $-2$ holds}}

\end{picture}

\vskip 5mm

The $(\al,\be)$ plane is divided into 4 regions by the lines $2\al=\pm\be$. Inside the right-hand region $Z(J)$ converges at both ends, inside the top and bottom regions $Z(J)$ converges only at, respectively, $2$ and $-2$, and inside the left-hand region $Z(J)$ diverges at both ends. As for the borderlines $2\al=\pm\be$, if $\al\ge 0$, then $Z(J)$ converges at both ends and if $\al<0$, then $Z(J)$ diverges at $\pm 2$ (convergence at $\mp 2$ is left open). The divergence results follow from \cite{SZ} and hold for more general errors, trace class in particular.

The author would like to thank Paul Nevai and Barry Simon for useful discussions.

\section{On an argument of Dombrowski-Nevai} \lb{S2}

In this section we will improve a result of Dombrowski-Nevai \cite{DN}. We will closely follow their presentation and introduce an additional twist which will yield this improvement. The notation here is slightly different from \cite{DN} because their $b_n$'s start with $n=0$ and their ``free'' $a_n$'s are $\tfrac 12$.
We define 
\begin{equation} \lb{2.1}
S_n(x)\equiv \sum_{j=0}^n \Bigl[ (a_{j+1}^2-a_j^2)P_j^2(x) + a_j(b_{j+1}-b_j)P_{j}(x)P_{j-1}(x) \Bigr]
\end{equation}
where we take $a_0=b_0=0$. Notice that the $S_n$ obey the obvious recurrence relation
\begin{equation} \lb{2.3}
S_{n}(x)=S_{n-1}(x) + (a_{n+1}^2-a_{n}^2)P_{n}^2(x) + a_{n}(b_{n+1}-b_{n})P_{n}(x)P_{n-1}(x) 
\end{equation}
Using this and \eqref{1.0a} one proves by induction the following formula from \cite{Dom}
\begin{equation} \lb{2.2}
S_n(x)=a_{n+1}^2 \biggl[ P_{n+1}^2(x) - \frac{x-b_{n+1}}{a_{n+1}}P_{n+1}(x)P_{n}(x) + P_{n}^2(x) \biggr]
\end{equation}

The results in \cite{DN} are based on \eqref{2.3} and \eqref{2.2}. Our simple but essential improvement is the introduction of a function closely related to $S_n$, but satisfying a recurrence relation which is more suitable for the purposes of this argument. We define
\begin{equation} \lb{2.4}
T_n(x)\equiv S_n(x) + \frac{a_{n+1}}2|b_{n+2}-b_{n+1}|P_n^2(x) 
\end{equation}
and then we have
\begin{align*}
T_{n}(x)= T_{n-1}(x) +& (a_{n+1}^2-a_{n}^2)P_{n}^2(x) + a_{n}(b_{n+1}-b_{n})P_{n}(x)P_{n-1}(x) \notag
\\ +& \frac{a_{n+1}}2|b_{n+2}-b_{n+1}|P_n^2(x) - \frac{a_{n}}2|b_{n+1}-b_{n}|P_{n-1}^2(x)
\end{align*}
The importance of this relation lies in the fact that it implies the crucial inequality
\begin{equation} \lb{2.6}
T_{n}(x)\le T_{n-1}(x) + \biggl[a_{n+1}^2-a_{n}^2 + \frac{a_{n+1}}2|b_{n+2}-b_{n+1}| + \frac{a_{n}}2|b_{n+1}-b_{n}| \biggr] P_n^2(x)
\end{equation}
by writing $|P_{n}(x)P_{n-1}(x)|\le \tfrac 12\bigl(P_{n}^2(x)+P_{n-1}^2(x)\bigr)$. Hence, our choice of $T_n$ eliminated the unpleasant cross term in \eqref{2.3}.

Now we are ready to apply the argument from \cite{DN}, but to $T_n$ in place of $S_n$. We define
\begin{equation} \lb{2.7}
\del_n\equiv \biggl[a_{n+1}^2-a_{n}^2 + \frac{a_{n+1}}2|b_{n+2}-b_{n+1}| + \frac{a_{n}}2|b_{n+1}-b_{n}|\biggr]_+
\end{equation}

\begin{lemma} \lb{L2.1}
If $a_{n}\ge 1+\frac{|b_n|}2$ for $n>N$, then for $n>N$
\begin{equation} \lb{2.8}
(4-x^2)P_n^2(x)\le 4T_{n-1}(x)   \qquad\quad |x|\le 2
\end{equation}
\begin{equation} \lb{2.9}
\max_{|x|\le 2}P_n^2(x)\le (n+1)^2\max_{|x|\le 2}T_{n-1}(x) 
\end{equation}
\begin{equation} \lb{2.10}
0\le T_n(x)\le  \exp\biggl( \frac{4\del_n}{4-x^2} \biggr) T_{n-1}(x)  \qquad\quad |x|< 2
\end{equation}
\begin{equation} \lb{2.11}
\max_{|x|\le 2}T_n(x)\le e^{(n+1)^2\del_n} \max_{|x|\le 2}T_{n-1}(x)
\end{equation}
\end{lemma}

\begin{proof}
From \eqref{2.2}
\[
S_{n-1}(x)=a_{n}^2 \biggl[ P_{n-1}(x)-\frac{x-b_{n}}{2a_{n}}P_{n}(x) \biggr]^2 + \frac 14 \biggl[ 4a_{n}^2-(x-b_n)^2 \biggr] P_{n}^2(x)
\]
The assumption $2a_n\ge 2+|b_n|$ implies $4a_{n}^2-(x-b_n)^2 \ge 4-x^2$ for $|x|\le 2$, and \eqref{2.4} implies $T_{n-1}(x)\ge S_{n-1}(x)$. This proves \eqref{2.8}. \eqref{2.9} follows from \eqref{2.8} and a theorem of Bernstein \cite[p.139]{Nat}, and \eqref{2.10}/\eqref{2.11} from \eqref{2.6}, \eqref{2.7} and \eqref{2.8}/\eqref{2.9}.
\end{proof}

In \cite{DN}, similar statements are proved for $S_n$. The important difference is that the proofs use \eqref{2.3} rather than \eqref{2.6}, and therefore involve $\del_n'=[a_{n+1}^2-a_{n}^2]_++a_{n}|b_{n+1}-b_{n}|$. This is a serious drawback because the condition $\sum n\del_n<\infty$ will play a central role in our considerations. If, for example, $a_n=1+\alpha/n$ and $b_n=\beta/n$, then $\sum n\del_n'<\infty$ only if $\alpha\ge 0$ and $\beta=0$ (cf.~the result from \cite{DN} mentioned in Section \ref{S1}), but $\sum n\del_n<\infty$ whenever $2\alpha\ge|\beta|$. This is because in $\del_n$ (and not in $\del_n'$) the contribution of the positive $|b_{n+1}-b_n|$ terms can be canceled by a decrease in $a_n$. Therefore $T_n$ can sometimes be a better object to look at than $S_n$, for example in the case of Coulomb Jacobi matrices.
The next result relates $T_n$ and $Z(J)$.

\begin{lemma} \lb{L2.2}
Suppose $\lim_{n\to\infty} a_n=1$, $\lim_{n\to\infty} b_n=0$ and 
\begin{equation} \lb{2.12}
\sum_{n=1}^\infty \big(|a_{n+1}-a_n|+|b_{n+1}-b_n|\big)<\infty
\end{equation}
Then for $|x|<2$
\begin{equation} \lb{2.13}
\lim_{n\to\infty} T_n(x)=\frac{\sqrt{4-x^2}}{2\pi \nu'(x)}
\end{equation}
\end{lemma}

{\it Remarks.}
1. The right-hand side appears in \eqref{1.1} and so one can use \eqref{2.13} and Fatou's lemma to obtain upper bounds on $Z(J)$ (see proof of Theorem \ref{T2.5}).
\smallskip

2. Results relating density of the absolutely continuous part of the spectral measure and asymptotics of the solutions of difference (or differential) equations, under the assumption of finite variation of the potential, go back to Weidmann \cite{Wei1,Wei2}.
\smallskip

\begin{proof}
If \eqref{2.12} holds, then it is proved in \cite{MN} that for $x\in(-2,2)$
\[
\lim_{n\to\infty}\biggl[ P_{n+1}^2(x) - \frac{x-b_{n+2}}{a_{n+2}}P_{n+1}(x)P_{n}(x) + \frac{a_{n+1}}{a_{n+2}}P_{n}^2(x) \biggr]
=\frac{\sqrt{4-x^2}}{2\pi \nu'(x)}
\]
(in \cite{MN} $a_n\to \frac 12$ and the limit is $2\sqrt{1-x^2}/\pi\nu'(x)$).
By \cite{Sim} $\{P_n(x)\}_n$ is bounded for any fixed $x\in(-2,2)$ when \eqref{2.12} holds. Hence $a_n\to 1$ and $b_n\to 0$ imply
\[
\lim_{n\to\infty}\bigl[ P_{n+1}^2(x) - xP_{n+1}(x)P_{n}(x) + P_{n}^2(x) \bigr]
=\frac{\sqrt{4-x^2}}{2\pi \nu'(x)}
\]
But by \eqref{2.2} and \eqref{2.4} this limit is the same as $\lim_n T_n(x)$.
\end{proof}

In the light of the discussion preceeding the lemma, the following will be useful.

\begin{lemma} \lb{L2.3}
If $\inf\{a_n\}>0$ and $\sum_n \del_n<\infty$, then \eqref{2.12} holds.
\end{lemma}

\begin{proof}
We have $0\le [a_{n+1}^2-a_n^2]_+\le \del_n$, hence $\sum [a_{n+1}^2-a_n^2]_+<\infty$. By telescoping $\sum [a_{n+1}^2-a_n^2]_-\le a_1^2+\sum [a_{n+1}^2-a_n^2]_+<\infty$ and so $\sum |a_{n+1}^2-a_n^2|<\infty$. Since $\inf\{a_n\}>0$, it follows that $\sum |a_{n+1}-a_n|<\infty$. Also, since
\[
0\le\frac{a_{n+1}}2|b_{n+2}-b_{n+1}| + \frac{a_{n}}2|b_{n+1}-b_{n}|\le \del_n+|a_{n+1}^2-a_{n}^2|
\]
and $a_n$ are bounded away from zero, $\sum |b_{n+1}-b_n|<\infty$.
\end{proof}

These lemmas have the same consequences as in \cite{DN}, but with $\del_n$ in place of $\del_n'$. Thus we can prove the following two results.

\begin{theorem} \lb{T2.4}
Suppose $a_n\ge 1+\frac{|b_n|}2$ for $n>N$, $\lim_{n\to\infty} a_n=1$ and 
\[
\sum_{n=1}^\infty n^2\biggl[a_{n+1}^2-a_{n}^2 + \frac{a_{n+1}}2|b_{n+2}-b_{n+1}| + \frac{a_{n}}2|b_{n+1}-b_{n}|\biggr]_+<\infty
\]
Then there is $c>0$ such that
\[
\frac{d\nu_{ac}}{dx}(x)\ge c\sqrt{4-x^2} \quad\quad |x|<2
\]
\end{theorem}

{\it Remarks.}
1. In particular, the corresponding matrix $J$ is Szeg\H o.
\smallskip

2. Notice that the above conditions are satisfied for $a_n\downarrow 1$, $b_n\equiv 0$, as pointed out in \cite{DN}.
\smallskip

\begin{proof}
By \eqref{2.7} and \eqref{2.11} we have for all $|x|\le 2$ and $n>N$
\[
T_n(x)\le \exp\Biggl( \sum_{j=N}^\infty (j+1)^2\del_j \Biggr) \max_{|x|\le 2} T_N(x)\equiv \frac 1{2\pi c}<\infty
\]
Lemmas \ref{L2.3} and \ref{L2.2} finish the proof. 
\end{proof}

The main result of this section is

\begin{theorem} \lb{T2.5}
Suppose $a_n\ge 1+\frac{|b_n|}2$ for $n>N$, $\lim_{n\to\infty} a_n=1$ and 
\[
\sum_{n=1}^\infty n\biggl[a_{n+1}^2-a_{n}^2 + \frac{a_{n+1}}2|b_{n+2}-b_{n+1}| + \frac{a_{n}}2|b_{n+1}-b_{n}|\biggr]_+<\infty
\]
Then $J$ given by \eqref{1.0} is Szeg\H o.
\end{theorem}

\begin{proof}
Once again, we closely follow \cite{DN}. By Lemmas \ref{L2.3} and \ref{L2.2} and Fatou's lemma
\[
Z(J)\le \lim_{\eps\downarrow 0} \biggl( \liminf_{n\to\infty} \frac 1{2\pi} \int_{-2+\eps}^{2-\eps} \ln_+(T_n(x))\,\frac{dx}{\sqrt{4-x^2}} \biggr)
\]
and so it is sufficient to prove
\[
\int_{0}^{2-n^{-2}} \ln_+(T_n(x))\,\frac{dx}{\sqrt{2-x}} + \int_{-2+n^{-2}}^0 \ln_+(T_n(x))\,\frac{dx}{\sqrt{2+x}} \le C
\]
for some $C<\infty$. Let us consider the first integral, which we denote $I_n$ (both can be treated similarly).

By \eqref{2.10} and \eqref{2.11}, for $n>N$ 
\begin{align*}
I_n \le & I_{n-1} + 2\del_n \int_0^{2-\frac 1{(n-1)^{2}}} \frac{dx}{(2-x)^{\frac 32}} 
+ \ln_+\biggl[\max_{|x|\le 2} T_n(x)\biggr] \int_{2-\frac 1{(n-1)^{2}}}^{2-\frac 1{n^{2}}} \frac{dx}{\sqrt{2-x}}
\\ = & I_{n-1} + 2\del_n(2n-2-\sqrt{2}) + \left(\frac 2{n-1} - \frac 2n\right) \ln_+\biggl[\max_{|x|\le 2} T_n(x)\biggr]
\\ \le & I_{n-1} + 4n\del_n + \frac 2{n-1} \ln_+\biggl[\max_{|x|\le 2} T_{n-1}(x)\biggr] 
+ \frac{2(n+1)^2\del_n }{n-1}- \frac 2n \ln_+\biggl[\max_{|x|\le 2} T_n(x)\biggr]
\\ \le & I_{n-1} + 13n\del_n + \frac 2{n-1} \ln_+\biggl[\max_{|x|\le 2} T_{n-1}(x)\biggr] 
- \frac 2n \ln_+\biggl[\max_{|x|\le 2} T_n(x)\biggr]
\end{align*}
because $\ln_+(x)+\ln_+(y)\ge \ln_+(xy)$. By iterating this, we obtain
\begin{align*}
I_n \le & I_N + 13\sum_{j=N+1}^n j\del_j + \frac 2N \ln_+\biggl[\max_{|x|\le 2} T_N(x)\biggr]
\\ \le & 13\sum_{n=1}^\infty n\del_n +  5\ln_+\biggl[\max_{|x|\le 2} T_N(x)\biggr] \equiv \frac C2
\end{align*}
as desired.
\end{proof}

In particular, if $a_n\equiv 1+\alpha/n$ and $b_n\equiv \beta/n$ with $2\alpha\ge|\beta|$, then $J$ is Szeg\H o. Later we will add $O(n^{-1-\eps})$ errors to these $a_n,b_n$.

For further reference we make

\begin{definition} \lb{D2.6}
We call a pair of sequences $\{a_n,b_n\}_{n=1}^\infty$ {\it admissible}, if $a_n\ge 1+\frac{|b_n|}2$ for $n>N$, $\lim_{n\to\infty} a_n=1$ and 
\[
\sum_{n=1}^\infty n\biggl[a_{n+1}^2-a_{n}^2 + \frac{a_{n+1}}2|b_{n+2}-b_{n+1}| + \frac{a_{n}}2|b_{n+1}-b_{n}|\biggr]_+<\infty
\]
\end{definition}

Hence, if $\{a_n,b_n\}$ is admissible, then $J$ is Szeg\H o. We make some useful observations.

\begin{lemma} \lb{L2.7}
Suppose $\{a_n,b_n\}$ is admissible and $\{e_n,f_n\}$ is such that $2e_n\ge|f_n|$ for $n>N$, $e_n\to 0$ and $\sum n\bigl( |e_{n+1}-e_n| + |f_{n+1}-f_n|\bigr)< \infty$. Then $\{a_n+e_n,b_n+f_n\}$ is also admissible.
\end{lemma}

\begin{proof}
We only need to show the last condition for admissibility. If 
\begin{align*}
\eps_n\equiv (a_{n+1}+e_{n+1})^2-(a_n+e_n)^2 & + \frac{a_{n+1}+e_{n+1}}2 |b_{n+2}+f_{n+2}-b_{n+1}-f_{n+1}| 
\\ & + \frac{a_n+e_n}2 |b_{n+1}+f_{n+1}-b_{n}-f_{n}|
\end{align*}
then we want $\sum n[\eps_n]_+<\infty$. Notice that
\begin{align*}
\eps_n \le \del_n & +2a_{n+1}|e_{n+1}-e_n| + 2|a_{n+1}-a_n||e_n| + |e_{n+1}+e_n||e_{n+1}-e_n|
\\ & + \frac{a_{n+1}+e_{n+1}}2 |f_{n+2}-f_{n+1}| + \frac{a_{n}+e_{n}}2 |f_{n+1}-f_{n}| 
\\ & + \frac{e_{n+1}}2 |b_{n+2}-b_{n+1}| + \frac{e_{n}}2 |b_{n+1}-b_{n}|
\end{align*}
and so we only need to prove $\sum nX_n<\infty$ for $X_n$ being any of the above terms. If $X_n$ is $\del_n$ or one of the terms containing $|e_{n+1}-e_n|$ or $|f_{n+1}-f_n|$, then this is obvious. For the remaining three terms the same is true by the fact that $\sum n|e_{n+1}-e_n|<\infty$ and $e_n\to 0$ imply $ne_n\to 0$, and by Lemma \ref{L2.3}.
\end{proof}

\begin{lemma} \lb{L2.8}
Suppose $\{a_n,b_n\}$ is admissible and $e_n\downarrow 0$ is such that $\{ne_n|a_{n+1}-a_n|\}$ or $\{ne_n|b_{n+2}-b_{n+1}|\}$ is bounded. Then $\{a_n+e_n,b_n\}$ is also admissible.
\end{lemma}

\begin{proof}
If
\begin{align*}
\eps_n\equiv (a_{n+1}+e_{n+1})^2-(a_n+e_n)^2 & + \frac{a_{n+1}+e_{n+1}}2 |b_{n+2}-b_{n+1}| 
\\ & + \frac{a_n+e_n}2 |b_{n+1}-b_{n}|
\end{align*}
then by $e_{n+1}\le e_n$
\begin{align*}
\eps_n \le & \del_n +2a_{n+1}e_{n+1} - 2a_ne_n + e_{n+1}^2-e_n^2 + \frac{e_{n+1}}2 |b_{n+2}-b_{n+1}| + \frac{e_{n}}2 |b_{n+1}-b_{n}|
\\ \le & \del_n+e_n \bigl( 2(a_{n+1}-a_n)+\tfrac 12|b_{n+2}-b_{n+1}| + \tfrac 12|b_{n+1}-b_{n}|  \bigr)
\\ \le & \del_n +\frac{2e_n}{a_{n+1}+a_n}\biggl( \del_n+\frac{|a_n-a_{n+1}|}4 |b_{n+2}-b_{n+1}|+ \frac{|a_{n+1}-a_n|}4 |b_{n+1}-b_n| \biggr)
\end{align*}
so $\sum n[\eps_n]_+<\infty$ by the hypotheses and Lemma \ref{L2.3}.
\end{proof}

We conclude this section with an interesting corollary. Notice that in \eqref{1.6} one would like to take $n\to\infty$ to pass from the step-by-step sum rule to a ``full size'' sum rule not involving $J^{(n)}$. For this, one would need to separate the terms in \eqref{1.6} when taking $n\to\infty$. The following shows that there are many Jacobi matrices which are Szeg\H o, but one cannot do this (see \cite{SZ} for results on when it is possible).

\begin{corollary} \lb{L2.9}
Let $\{a_n,b_n\}$ be admissible and let $\til J$ be a matrix with $\til a_n\equiv a_n+c/n$ and $\til b_n\equiv b_n$ for some $c>0$. Then $Z(\til J)<\infty$ but
\[
\bar A_0(\til J)\equiv\limsup_n \biggl(-\sum_{j=1}^n \ln(\til a_j)\biggr) =-\infty 
\]
and
\[
\calE_0(\til J)\equiv\sum_{j,\pm}\ln|\beta_j^\pm(\til J)|=\infty
\]
\end{corollary}

\begin{proof}
$\til J$ is Szeg\H o by Lemma \ref{L2.8}. Since $Z(J)<\infty$, \eqref{1.6} yields $\bar A_0(J)<\infty$ (because the other two terms in \eqref{1.6} are bounded from below). Since $a_n\to 1$ and $\sum \tfrac cn=\infty$, we obtain $\bar A_0(\til J)=-\infty$. By Theorem 4.1(d) in \cite{SZ}, this implies $\calE_0(\til J)=\infty$.
\end{proof}

\section{Control of change of eigenvalues under perturbations} \lb{S3}

In this section we will prove results on the behavior of eigenvalues under certain finite-rank perturbations of the $a_n$'s and $b_n$'s. Namely, we will show that these perturbations decrease $E_j^+$ and increase $E_j^-$ for all but finitely many $j$.  This, of course, means that we will not consider arbitrary perturbations. Indeed, in all the perturbations we can treat, the $a_n$'s cannot increase. Immediately a question arises, how is this compatible with the possibility of $a_n>c_n$ in Theorem \ref{T1.2}. The answer is in Lemma \ref{L2.7}. Before doing a general $O(n^{-1-\eps})$ perturbation of $c_n,d_n$, we will increase the $c_n$'s by $Cn^{-1-\eps}$ for some large $C$, so that the assumptions of Theorem \ref{T1.2} will stay valid and the new $c_n$ will be larger than $a_n$. Then we will use results from this section. For details see the proof of Theorem \ref{T4.3}. 

For $j\ge 1$ and $n\ge -1$ we define
\[
p_n(\pm j)\equiv \frac{P_n(E_j^\pm)}{\Bigl( \sum_{m=0}^\infty {\displaystyle P_m^2}\bigl( E_j^\pm \bigr) \Bigr)^{\frac 12}}
\]
Hence $p(\pm j)\equiv \{ p_n(\pm j)\}_{n=0}^\infty$ is the normalized eigenfunction for energy $E_j^\pm$. Therefore $p(\pm j)$ satisfies the same recurrence relation as $P(E_j^\pm)$, and so
\begin{equation} \lb{3.1}
p_{n+1}(\pm j)= \frac{E_j^\pm-b_{n+1}}{a_{n+1}} p_n(\pm j) - \frac{a_n}{a_{n+1}} p_{n-1}(\pm j)
\end{equation}

In what follows, we will use the following result from first order perturbation theory for eigenvalues (see, e.g., \cite[p.151]{Thi}).

\begin{proposition} \lb{P3.1} 
Let $J(t)\equiv J+tA$ for $t\in(-\eps,\eps)$ where $J$ and $A$ are bounded self-adjoint operators on a Hilbert space. Assume that $J(0)$ has a simple isolated eigenvalue $E(0)\notin \sigma_{ess}(J(0))$ and let $\varphi(0)$ be the corresponding normalized eigenfunction. Then there are analytic functions $E(t)$, $\varphi(t)$ defined on some interval $(-\eps',\eps')$ such that $E(t)$ is a simple isolated eigenvalue of $J(t)$ with normalized eigenfunction $\varphi(t)$, and we have $\partt E(t)=\langle \varphi(t),A\varphi(t) \rangle$.
\end{proposition}

In the case of Jacobi matrices, all eigenvalues outside $[-2,2]$ are simple. Hence if $J(t)\equiv J+tA$ with $A$ bounded self-adjoint matrix, then 
\begin{equation} \lb{3.3}
\partt E_j^\pm(t) = \lan p(\pm j;t), A p(\pm j;t) \ran
\end{equation}
as long as $E_j^\pm(t)$ stays outside $[-2,2]$.

We define $E_j^\pm\equiv\pm 2$ whenever $J$ has less than $j$ positive/negative eigenvalues. Then, of course, \eqref{3.3} does not apply when $E_j^\pm(t)=\pm 2$, but we at least have continuity of $E_j^\pm(t)$ in $t$ by norm-continuity of $J(t)$. 

Here is the main idea of this section. Fix $n$ and take $A$ to be the matrix with $A_{n-1,n}=A_{n,n-1}=-1$ and all other entries zero (the upper left-hand corner of $A$ being $A_{0,0}$). Then increasing $t$ corresponds to decreasing $a_n$. We have 
\[
\partt E_j^\pm(t)=-2p_{n}(\pm j;t)p_{n-1}(\pm j;t)
\]
Let us take $j=1$. Then by the Sturm oscillation theory \cite{Tes} we know that $\sgn(p_n(1;t))=\sgn(p_{n-1}(1;t))$ and $\sgn(p_n(-1;t))=-\sgn(p_{n-1}(-1;t))$ for $n\ge 1$. Hence $E_1^+$ will decrease and $E_1^-$ will increase when we decrease $a_n$. This is exactly what we want.

Unfortunately, this is not always the case for other eigenvalues. Indeed, let us consider a positive eigenvalue $E_j^+$. By oscillation theory, $p(j)$ changes sign $j-1$ times and so $E_j^+$ will grow at certain $n$. However, if $E_j^+\approx 2$, $a_n\approx 1$ and $b_n\approx 0$, then by \eqref{3.1} $p_{n+1}(j)\approx 2p_n(j)-p_{n-1}(j)$, that is, $p(j)$ is (locally) close to a linear function of $n$. Therefore, if $\sgn(p_n(j))=-\sgn(p_{n-1}(j))$, then $\sgn(p_m(j))=\sgn(p_{m-1}(j))$ for $m\neq n$ but close to $n$. Hence, a suitable decrease of $a_n$ along with some neighboring $a_m$'s should always result into a decrease of $E_j^+$. This is the content of the present section.

\begin{definition} \lb{D3.2}
Let $\del>0$. We say that $\til J$ {\it $\del$-minorates\/} $J$, if $|E_j^\pm(\til J)|\le |E_j^\pm(J)|$ whenever $|E_j^\pm(J)|<2+\del$.
\end{definition}

{\it Remarks.}
1. This is well defined because $E_j^\pm\equiv \pm 2$ whenever $J$ has less than $j$ positive/negative eigenvalues.
\smallskip

2. Notice that for fixed $\del$ this relation is transitive.
\smallskip

\begin{lemma} \lb{L3.3}
There exists $\del>0$ such that the following is true. If for some $J$ we have $|a_m-1|<\del$ and $|b_m|<\del$ for $m\in\{n,n+1,n+2\}$, and $\til J$ is obtained from $J$ by decreasing $a_n$ by $c>0$ and $a_{n+2}$ by $d>0$ so that $|a_n-c-1|<\del$, $|a_{n+2}-d-1|<\del$ and $c/d\in [\tfrac 1{13},13]$, then $\til J$ $\del$-minorates $J$.
\end{lemma}

{\it Remark.}
That is, decreasing both $a_n$ and $a_{n+2}$ results into decrease of all but finitely many $|E_j^\pm|$. The same trick applied to $a_n$ and $a_{n+1}$ fails.
\smallskip

\begin{proof}
Let $q\equiv c/d$. Let $E\equiv E_j^+$ and $p_n\equiv p_n(+j)$ for some $2<E_j^+<2+\del$. Then by \eqref{3.1}
\begin{align}  \lb{3.4}
p_{n+1}=& 2p_n-p_{n-1} + \frac{E-2a_{n+1}-b_{n+1}}{1+(a_{n+1}-1)}p_n + \frac{a_{n+1}-a_n}{1+(a_n-1)}p_{n-1} \notag
\\ =& 2p_n-p_{n-1} + O(\del)(|p_n|+|p_{n-1}|) 
\end{align}
with $|O(\del)|\le C\del$ for some universal $C<\infty$ and all small $\del$.
Similarly we obtain by iterating \eqref{3.1}
\begin{equation} \lb{3.5}
p_{n+2}=3p_n-2p_{n-1} + O(\del)(|p_n|+|p_{n-1}|)
\end{equation} 

Let now $J(t)\equiv J+tA$ where $A$ is such that $A_{n-1,n}=A_{n,n-1}=-q$, $A_{n+1,n+2}=A_{n+2,n+1}=-1$ and all other entries are $0$. Then obviously $E_j^\pm(0)=E_j^\pm$ and $\til J=J(d)$. By \eqref{3.3}
\[
\partt E_j^+(0)=\lan p,Ap\ran=-2(qp_{n}p_{n-1}+p_{n+2}p_{n+1})
\]
By \eqref{3.4} and \eqref{3.5}
\begin{equation} \lb{3.6}
qp_{n}p_{n-1}+p_{n+2}p_{n+1}=6p_n^2-(7-q)p_np_{n-1}+2p_{n-1}^2+O(\del)(p_n^2+p_{n-1}^2)
\end{equation} 
Since $6\cdot 2-\left(\frac{7-q}2\right)^2>0$ for $q\in(7-4\sqrt{3},7+4\sqrt{3})\supset[\tfrac 1{13},13]$, it follows that \[
6p_n^2-(7-q)p_np_{n-1}+2p_{n-1}^2>|O(\del)|(p_n^2+p_{n-1}^2)
\]
for small enough $\del$ (uniformly for all $q\in [\tfrac 1{13},13]$). That is, $\partt E_j^+(0)<0$.

This argument obviously applies to all $t\in[0,d]$, not only to $t=0$, as long as $E_j^+(t)>2$. This is because for each such $t$, $J(t)$ satisfies the conditions of this lemma. Hence $E_j^+(t)$ can only decrease with $t$ (and so stays smaller than $2+\del$). Also, no new eigenvalues can appear. Indeed -- if $E_j^+(t_1)=2$ and $E_j^+(t_2)>2$ for some $t_2>t_1$, then $E_j^+(t)$ would have to have a discontinuity in $[t_1,t_2]$, because by the above argument it has to decrease whenever it is larger than 2.

A similar argument applies to $E_j^-(0)>-2-\del$, with $p_{n+1}\approx -2p_n-p_{n-1}$ and $p_{n+2}\approx 3p_n+2p_{n-1}$ in place of \eqref{3.4} and \eqref{3.5}, and shows that such $E_j^-$ increases with $t$. The result follows.
\end{proof}

As mentioned earlier, same trick with $a_{n+1}$ in place of $a_{n+2}$ does not work. Indeed -- in \eqref{3.6} we would have
$2p_n^2-(1-q)p_np_{n-1}+O(\del)(p_n^2+p_{n-1}^2)$ which cannot be guaranteed to be positive for any $\del>0$. However, we can replace $a_{n+2}$ by $a_{n+k}$ for $k\ge 2$, and the lemma stays valid for some smaller $\del=\del(k)>0$ and $c/d\in[(4k^2-3)^{-1},4k^2-3]$ (we use that $p_{n+k}\approx (k+1)p_n-kp_{n-1}$). Of course, the bounds on $|a_m-1|$ and $|b_m|$ have to hold for $m\in\{n,\dots,n+k\}$.

Before we start perturbing the $b_n$'s, let us state one more result with the same flavor.

\begin{lemma} \lb{L3.4}
There exists $\del>0$ such that the following is true. If for some $J$ we have $|a_m-1|<\del$ and $|b_m|<\del$ for $m\in\{n,n+1,n+2\}$, and $\til J$ is obtained from $J$ by decreasing $a_n$, $a_{n+1}$ and $a_{n+2}$ by $c>0$ so that $|a_m-c-1|<\del$ for $m\in\{n,n+1,n+2\}$, then $\til J$ $\del$-minorates $J$.
\end{lemma}

 {\it Remark.}
Again, the result can be extended to decreasing $a_n,\dots,a_{n+k}$ (for $k\ge 2$) by $c>0$, with a smaller $\del=\del(k)>0$.
\smallskip

\begin{proof}
An argument as above yields for $A_{n-1,n}=A_{n,n-1}=A_{n,n+1}=A_{n+1,n}=A_{n+1,n+2}=A_{n+2,n+1}=-1$ 
\begin{align*}
\partt E_j^+(0) & =-2(p_{n-1}p_n+p_np_{n+1}+p_{n+1}p_{n+2})
\\ & =-2(8p_n^2-7p_np_{n-1}+2p_{n-1}^2+O(\del)(p_n^2+p_{n-1}^2))
\end{align*}
which is negative for small enough $\del$, since $8\cdot 2-(\tfrac 72)^2>0$. The rest of the previous proof applies.
\end{proof}

Our next aim is to allow perturbations of the $b_n$'s as well. If one decreases $b_n$, it is obvious that all $E_j^+$ decrease, but all $E_j^-$ decrease as well. Hence, perturbing the $b_n$'s alone will not move ``in'' all eigenvalues. To ensure that, we have to counter the undesired movement of $E_j^-$ by decreasing $a_n$'s.

\begin{lemma} \lb{L3.5}
There exists $\del>0$ such that the following is true. If for some $J$ we have $|a_m-1|<\del$ and $|b_m|<\del$ for $m\in\{n,n+1,n+2\}$, and $\til J$ is obtained from $J$ by decreasing $a_n$ and $a_{n+2}$ by $c>0$ and changing $b_n$ by $d\in[-\tfrac c2,\tfrac c2]$ so that $|a_n-c-1|<\del$, $|a_{n+2}-c-1|<\del$ and $|b_n+d|<\del$, then $\til J$ $\del$-minorates $J$.
\end{lemma}

\begin{proof}
This time we have $A_{n-1,n}=A_{n,n-1}=A_{n+1,n+2}=A_{n+2,n+1}=-1$ and $A_{n-1,n-1}=q\equiv d/c$. We obtain
\begin{align*}
\partt E_j^+(0) & =-2(p_{n-1}p_n+p_{n+1}p_{n+2})+qp_{n-1}^2
\\ & =-2(6p_n^2-6p_np_{n-1}+(2-\tfrac q2)p_{n-1}^2+O(\del)(p_n^2+p_{n-1}^2))
\end{align*}
which is negative for small enough $\del$ if $q<1$ (i.e.~if $6\cdot (2-\tfrac q2)-(\tfrac 62)^2>0$). A similar argument for $E_j^-$ requires $q>-1$, so there is a  $\del>0$ which works for all $q\in[-\tfrac 12,\tfrac 12]$.
\end{proof}

\section{The main result} \lb{S4}

We will now outline an argument suggested in \cite{SZ}. This shows how to use \eqref{1.6} to prove stability of the Szeg\H o condition under certain trace class perturbations. 

Let $\til J$ be a trace class perturbation of a matrix $J$ which we know to be Szeg\H o. That is
\begin{equation} \lb{4.00}
\sum_n\bigl(|\til a_n-a_n|+|\til b_n-b_n|\bigr)<\infty
\end{equation}
Let $\til J_n$ be the matrix which we obtain from $J$ by replacing $a_j,b_j$ by $\til a_j,\til b_j$ for $j=1,\dots,n$. Then $\til J_n\to \til J$ pointwise (and also in norm). Now by applying \eqref{1.6} to both $\til J_n$ and $J$ and subtracting, we obtain
\begin{equation} \lb{4.0a}
Z(\til J_n) = Z(J) -\sum_{j=1}^n (\ln (\til a_j)-\ln (a_j)) + \sum_{j,\pm}\Big(\ln|\be_j^\pm(\til J_n)| - \ln|\be_j^\pm(J)|\Big)
\end{equation}
By lower semicontinuity of $Z$ in $J$ (in the topology of pointwise convergence of matrix elements; see \cite{KS}), we know that $Z(\til J)\le \liminf Z(\til J_n)$. So taking $n\to\infty$ in \eqref{4.0a} we obtain
\begin{equation} \lb{4.0b}
Z(\til J) \le Z(J) +\sum_{j=1}^\infty |\ln (\til a_j)-\ln (a_j)| + \liminf_{n}\sum_{j,\pm}\Big(\ln|\be_j^\pm(\til J_n)| -\ln|\be_j^\pm(J)|\Big)
\end{equation}

If $\inf_j\{\til a_j,a_j\}>0$, then the first sum is finite by \eqref{4.00}.
Hence, if we could show that the $\liminf$ is smaller than $+\infty$, we would prove $\til J$ to be Szeg\H o. Notice that this is true if for some $\del>0$ each $\til J_n$ $\del$-minorates $J$, because then $|\be_j^\pm(\til J_n)|\le |\be_j^\pm(J)|$ whenever $|E_j^\pm(J)|<2+\del$ and the other $|\be_j^\pm(\til J_n)|$ are bounded. This is where results from the previous section enter the picture.

Unfortunately, we cannot treat general trace class perturbations at this moment. The reason is the necessity to use Lemma \ref{L2.7}, as described in Section \ref{S3}. It also needs to be said  that in what follows, the ``partial perturbations'' $\til J_n$ will be slightly different from those above. They will differ in up to 4 matrix elements, but they will still converge to $\til J$ and so \eqref{4.0b} will stay valid. 

Let us now apply the above argument. We start with

\begin{lemma} \lb{L4.1}
Let $J$ be Szeg\H o with $a_n\to 1$, $b_n\to 0$, and let $e_n\downarrow 0$, $e_n<a_n$, $\sum_n e_n<\infty$. Then the matrix $\til J$ with $\til a_n\equiv a_n-e_n$ and $\til b_n\equiv b_n$ is also Szeg\H o.
\end{lemma}

\begin{proof}
Let $\del\equiv \min\{\del(2),\del(3),\del(4)\}>0$ where $\del(k)$ are as in the remark after Lemma \ref{L3.4} (that is, good for decreasing 3, 4 and 5 consecutive $a_n$'s). Let $N$ be such that for $j\ge N$ we have $|a_j-1|<\del$, $|\til a_j-1|<\del$ and $|b_j|<\del$. 
For $n\ge N+1$ let $\til J_n$ be such that $b_j(\til J_n)\equiv b_j$ and 
\[
a_j(\til J_n)\equiv \begin{cases} \til a_j & j\le N-1 \\ \til a_j+e_{n+1} & N\le j\le n   \\ a_j & j\ge n+1 \end{cases}
\]
Then $\til J_{N+1}$ is Szeg\H o because it is a finite-rank perturbation of $J$. 

Let $n\ge N+2$. Notice that $\til J_{n}$ is obtained from $\til J_{n-1}$ by decreasing $a_j(\til J_{n-1})$ by $c\equiv e_n-e_{n+1}$ for $j=N,\dots,n$. This can be accomplished by successive decreases of 3, 4 or 5 neighboring $a_j$'s by $c$, as in Lemma \ref{L3.4} (and the remark after it). It follows that $\til J_n$ $\del$-minorates $\til J_{n-1}$, and so by induction $\til J_n$ $\del$-minorates $\til J_{N+1}$.
Then by \eqref{4.0a} (with $\del<\tfrac 12$)
\[
Z(\til J_n)\le Z(\til J_{N+1})+2\sum_{j=N}^\infty e_j+ K\ln(M)<\infty
\]
where $K$ is the number of eigenvalues of $\til J_{N+1}$ outside $(-2-\del,2+\del)$ and $M\equiv 3\sup_j\{a_j,|b_j|\}\ge\|\til J_n\|$. So $Z(\til J_n)$ are uniformly bounded and since $\til J_n\to \til J$ pointwise, lower semicontinuity of $Z$ implies $Z(\til J)<\infty$.
\end{proof}

\begin{corollary} \lb{C4.2}
Suppose $\{a_n,b_n\}$ is admissible and $\{e_n,f_n\}$ is such that $e_n\to 0$, $f_n\to 0$, $e_n> -a_n$ and $\sum n\bigl( |e_{n+1}-e_n| + |f_{n+1}-f_n|\bigr)< \infty$. Then the matrix $\til J$ with $\til a_n\equiv a_n+e_n$, $\til b_n\equiv b_n+f_n$ is Szeg\H o.
\end{corollary}

{\it Remark.}
This is almost like Lemma \ref{L2.7} with the condition $2e_n\ge |f_n|$ removed.
\smallskip

\begin{proof}
Let us define $\bar e_n\equiv \sum_{j=n}^\infty|e_{j+1}-e_j|$ and similarly for $f_n$. Notice that $\bar e_n\ge |e_n|$, $\bar e_n\downarrow 0$ and
\[
\sum_{n=1}^\infty \bar e_n\le\sum_{n=1}^\infty n|e_{n+1}-e_n|<\infty
\]
Then if $\til e_n\equiv e_n+\bar e_n+\bar f_n$, we have $2\til e_n\ge |f_n|$, and so $\{a_n+\til e_n,b_n+f_n\}$ is admissible by Lemma \ref{L2.7}. Then by Lemma \ref{L4.1} the result follows.
\end{proof}

We are now ready to prove Theorem \ref{T1.2}.

\begin{theorem} \lb{T4.3}
Suppose that $\{a_n,b_n\}$ is admissible and $\eps>0$. Then the matrix $\til J$ with
\[
\til a_n\equiv a_n + O(n^{-1-\eps}) \qquad \qquad \til b_n\equiv b_n +O(n^{-1-\eps})
\]
is Szeg\H o.
\end{theorem}

\begin{proof}
Our strategy is as outlined in Section \ref{S3}. We let 
\begin{equation} \lb{4.0}
C\equiv \sup_n\{|\til a_n-a_n|n^{1+\eps}, |\til b_n-b_n|n^{1+\eps}\}<\infty
\end{equation}
and increase $a_n$ by $6Cn^{-1-\eps}$ (we call these again $a_n$). Then by Lemma \ref{L2.7} (or Lemma \ref{L2.8}), $\{a_n,b_n\}$ (with the new $a_n$) is also admissible. Thus, the new $J$ is Szeg\H o and we now have 
\begin{align} \lb{4.1}
a_n-\til a_n  & \in [5Cn^{-1-\eps}, 7Cn^{-1-\eps}] 
\\ b_n-\til b_n & \in [-Cn^{-1-\eps}, Cn^{-1-\eps}] \notag
\end{align}

Let $\del$ be such that both Lemmas \ref{L3.3} and \ref{L3.5} hold. Let $N$ be such that for $j\ge N$ we have $|a_j-1|<\del$, $|\til a_j-1|<\del$, $|b_j|<\del$ and $|\til b_j|<\del$. 
We let $\til J_{N-1}$ be such that 
\[
a_j(\til J_{N-1})\equiv \begin{cases} \til a_j & j\le N-1 \\ a_j & j\ge N \end{cases}
\]
and similarly for $b_j(\til J_{N-1})$. Then $\til J_{N-1}$ is Szeg\H o because it is a finite-rank perturbation of $J$. 

We construct $\til J_N$ from $\til J_{N-1}$ by first decreasing $a_N,a_{N+2}$ by $2|b_N-\til b_N|$ and changing $b_N$ to $\til b_N$, and then decreasing $a_N$ by $a_N-\til a_N$ and $a_{N+2}$ by $(a_N-\til a_N)/13$ (in terms of the new $a_N$). Both these perturbations are $\del$-minorating by Lemmas \ref{L3.3} and \ref{L3.5}, and the obtained matrix $\til J_N$ agrees with $\til J$ in first $N$ couples $a_j(\til J_N)$, $b_j(\til J_N)$. The others are same as in $J$, only exception being $a_{N+2}(\til J_N)$, for which we know
\begin{equation} \lb{4.2}
a_{N+2}(\til J_{N})-\til a_{N+2} \in [2C(N+2)^{-1-\eps}, 7C(N+2)^{-1-\eps}] 
\end{equation}
(if $N$ is chosen so that $33N^{-1-\eps}\le 39(N+2)^{-1-\eps}$).

Now we apply the same procedure to inductively construct $\til J_n$ from $\til J_{n-1}$ for $n\ge N+1$. Each $\til J_n$ will agree with $\til J$ up to index $n$, and other elements will be the same as in $J$, with the exception of $a_{n+1}(\til J_n)$ and $a_{n+2}(\til J_n)$. For these we will have \eqref{4.2} (with $n+1$ and $n+2$ in place of $N+2$), which is just enough so that we can change $b_{n+1}$ to $\til b_{n+1}$ when passing to $\til J_{n+1}$ by the same method. Since $\til J_n$ $\del$-minorates $\til J_{n-1}$, we obtain by induction that each $\til J_n$ $\del$-minorates $\til J_{N-1}$.

Again, we have by \eqref{4.0a} (with $\del<\tfrac 12$)
\[
Z(\til J_n)\le Z(\til J_{N-1})+14C\sum_{j=N}^\infty j^{-1-\eps}+ K\ln(M)<\infty
\]
with $K$ and $M$ as in the proof of Lemma \ref{L4.1}. Since $\til J_n\to \til J$, the result follows.
\end{proof}

\begin{corollary} \lb{C4.4}
Let $2\al\ge|\be|$, $e_n\downarrow 0$, $\eps>0$ and
\[
a_n\equiv 1+\frac \al n +e_n+O(n^{-1-\eps}) \qquad \qquad b_n\equiv \frac \be n +O(n^{-1-\eps})
\]
Then $J$ is Szeg\H o.
\end{corollary}

{\it Remarks.}
1. This settles the $2\al\ge|\be|$ case of Askey's conjecture.
\smallskip

2. The same is true when $\al n^{-1}$, $\be n^{-1}$ are replaced by $\al n^{-\gamma}$, $\be n^{-\gamma}$ for any $\gamma>0$.
\smallskip

\begin{proof}
By Lemma \ref{L2.8}, $\{1+\al/n+e_n,\be/n\}$ is admissible. Then use Theorem \ref{T4.3}.
\end{proof}

Let us now return to considering perturbations of a single $a_n$. As noted in Section \ref{S1}, decreasing it can only guarantee decrease of $|E_1^\pm|$. However, if we know that $J$ has {\it no\/} bound states (eigenvalues outside $[-2,2]$), then this is sufficient to conclude that no new bound states can appear when decreasing $a_n$. 

\begin{theorem} \lb{T4.5}
Assume that $J$ with $a_n\to 1$, $b_n\to 0$ has only finitely many bound states and let $\til J$ have $\til a_n\le a_n$ and $\til b_n= b_n$ with $\til a_n\to 1$. Then $\til J$ is Szeg\H o if and only if $J$ is Szeg\H o and $\sum_n (a_n-\til a_n)<\infty$. In any case, $\til J$ also has only finitely many bound states.
\end{theorem}

\begin{proof}
We only need to prove this theorem for $J$ with no bound states. For by Sturm oscillation theory, $J$ has finitely many of them iff $J^{(n)}$ has none for large enough $n$. And $J$ is Szeg\H o iff $J^{(n)}$ is. So let us assume that $J$ has no bound states. Then by the above discussion, $\til J$ has none as well. Indeed -- if we let $\til J_n$ have $a_j(\til J_n)\equiv \til a_j$ for $j=1,\dots,n$ and all other entries same as $J$, then $\til J_n$ is created from $\til J_{n-1}$ by decreasing $a_n$. Since $\til J_{n-1}$ has no bound states, the same must be true for $\til J_{n}$. Since $\til J_n\to\til J$ in norm, $\til J$ also has no bound states.

If $Z(J)<\infty$ and $\sum(a_n-\til a_n)<\infty$, then $Z(\til J)<\infty$ by \eqref{4.0b}. No bound states and Theorem 4.1(d) in \cite{SZ} imply $\bar A_0(J)>-\infty$. So if $\sum(a_n-\til a_n)=\infty$, we obtain $\bar A_0(\til J)=\infty$, and then $Z(\til J)=\infty$ by \eqref{1.6} (since $Z(\til J^{(n)})\ge -\frac 12 \ln(2)$). Finally, if $Z(J)=\infty$, then no bound states and Theorem 4.1(a) in \cite{SZ} give $\bar A_0(J)=\infty$. This implies $\bar A_0(\til J)=\infty$ and so again $Z(\til J)=\infty$.
\end{proof}

Since Theorem 4.1 in \cite{SZ} does not distinguish between no bound states and $\calE_0(J)<\infty$, we can extend the above result to that case, but we need to restrict it to $\del$-minorating perturbations of the $a_n$'s only (e.g.~decreasing $a_n$ by $e_n\downarrow 0$). If $\calE_0(J)=\infty$, then such a result cannot be generally true. For example, if $2\al>|\be|$ in the Coulomb case, then decreasing $\al$ by $\al-|\be|/2$ results into a non-summable change of the $a_n$'s, but the matrix stays Szeg\H o.

\section{One-sided Szeg\H o conditions} \lb{S5}

In this section we will discuss Jacobi matrices which are Szeg\H o at 2 or $-2$. That is such, for which the Szeg\H o integral \eqref{1.1} converges at $\pm 2$, but is allowed to diverge at $\mp 2$. This is particularly interesting for $J$ which are Hilbert-Schmidt (i.e.~$L^2$) perturbations of $J_0$. For such $J$ we know from \cite{KS} that
\begin{equation} \lb{5.1}
Z_2^-(J)\equiv \f{1}{4\pi} \int_{-2}^2 \ln \biggl( \f{\sqrt{4-x^2\,}}{2\pi \nu'(x)}\biggr) \sqrt{4-x^2} \, dx <\infty
\end{equation}
(and $Z_2^-(J)\ge 0$ holds always; see \cite{KS}). That of course means that $Z(J)$ can only diverge at $\pm 2$.

We define
\begin{equation} \lb{5.2}
Z_1^\pm(J)\equiv \f{1}{4\pi} \int_{-2}^2 \ln \biggl( \f{\sqrt{4-x^2\,}}{2\pi \nu'(x)}\biggr) \f{2\pm x}{\sqrt{4-x^2}} \, dx 
\end{equation}
(the notation in \eqref{5.1} and \eqref{5.2} is from \cite{SZ}).
Again, $Z_1^\pm(J)$ is bounded below by some $c_0>-\infty$ and it is lower semicontinuous in $J$ \cite{SZ}. If $J-J_0\in L^2$, then by \eqref{5.1}, the integral \eqref{1.1} converges at $\pm 2$ if and only if $Z_1^\pm(J)<\infty$. Since we will mainly be interested in $J-J_0$ Hilbert-Schmidt, we use the following definition of one-sided Szeg\H o conditions from \cite{SZ}.

\begin{definition} \lb{D5.1}
We say that $J$ is {\it Szeg\H o at $\pm 2$\/} iff $Z_1^\pm(J)<\infty$.
\end{definition}

We consider $Z_2^-$ and $Z_1^\pm$ as above because they appear in sum rules similar to \eqref{1.6} \cite{SZ}.
Here we will only use the $Z_1^\pm$ sum rules. If we let $\xi^\pm(\beta)\equiv \ln|\beta|\pm \tfrac 12(\beta-\beta^{-1})$, then \cite{SZ} proves for $J-J_0$ compact
\begin{align} \lb{5.3}
Z_1^+(J) = & -\sum_{j=1}^n \bigl(\ln (a_j) + \tfrac 12 b_j \bigr) +\sum_{j,\pm}\biggl[ \xi^+\bigl(\beta_j^\pm(J)\bigr) -\xi^+\bigl(\beta_j^\pm(J^{(n)})\bigr) \biggr] + Z_1^+(J^{(n)})
\\ Z_1^-(J) = & -\sum_{j=1}^n \bigl(\ln (a_j) - \tfrac 12b_j \bigr) +\sum_{j,\pm}\biggl[ \xi^-\bigl(\beta_j^\pm(J)\bigr) - \xi^-\bigl(\beta_j^\pm(J^{(n)})\bigr) \biggr] + Z_1^-(J^{(n)})\notag
\end{align}
Just as with $Z(J)$, the infinite sums are always absolutely convergent and \eqref{5.3} holds even if $Z_1^\pm(J)=\infty$. This shows that the one-sided Szeg\H o conditions are also stable under finite-rank perturbations.

We will only consider the Szeg\H o condition at 2 and use the first of these identities. The reason for this is an obvious symmetry -- a matrix $J$ is Szeg\H o at $-2$ iff $\til J$ with $\til a_n\equiv a_n$ and $\til b_n\equiv -b_n$ is Szeg\H o at 2 (because $J\cong -\til J$). Therefore, our results for $+2$ will immediately translate into similar results for $-2$.

The main tool for handling trace class perturbations will be the following inequality, which we obtain from the first equation in \eqref{5.3} just as we obtained \eqref{4.0b} from \eqref{1.6} (with the same $\til J_n$).
\begin{align} \lb{5.4}
Z_1^+(\til J) \le Z_1^+(J) & +\sum_{j=1}^\infty |\ln (\til a_j)-\ln (a_j)|+\tfrac 12\sum_{j=1}^\infty |\til b_j-b_j|
\\ & + \liminf_{n}\sum_{j,\pm}\Big(\xi^+(\be_j^\pm(\til J_n)) -\xi^+(\be_j^\pm(J))\Big)  \notag
\end{align}
Notice that $\xi^+(\beta)$ is increasing and positive on $[1,\infty)$, and increasing and negative on $(-\infty,-1]$. That of course means that the last sum in \eqref{5.4} will be negative whenever $\beta_j^\pm(\til J_n)\le \beta_j^\pm(J)$ for all $j,\pm$. In particular, if $\til a_j=a_j$ and $\til b_j\le b_j$ for all $j$.

\begin{theorem} \lb{T5.2} Suppose $J-J_0$ is compact.

\begin{SL} 
\item[{\rm{(i)}}] If $J$ is Szeg\H o at $2$, and $\til J$ has $\til a_n= a_n$, $\til b_n\le b_n$ with $\sum_n(b_n-\til b_n)<\infty$, then $\til J$ is also Szeg\H o at $2$.
\item[{\rm{(ii)}}]  If $J$ is Szeg\H o at $-2$, and $\til J$ has $\til a_n= a_n$, $\til b_n\ge b_n$ with $\sum_n(\til b_n-b_n)<\infty$, then $\til J$ is also Szeg\H o at $-2$.
\item[{\rm{(iii)}}] Let $\hat J$ have $\hat a_n= a_n$, $\hat b_n\ge b_n$ with $\sum_n(\hat b_n- b_n)<\infty$, and let both $J, \hat J$ be Szeg\H o. If $\til J$ has $\til a_n= a_n$ and $b_n\le \til b_n\le \hat b_n$, then $\til J$ is also Szeg\H o.
\end{SL}
\end{theorem}

\begin{proof}
(i) follows from the discussion above, (ii) from (i) by symmetry, and (iii) from (i) and (ii) and the fact that $J$ is Szeg\H o iff it is Szeg\H o at both $\pm 2$.
\end{proof}

When perturbing the $a_n$'s as in Section \ref{S3}, we have to be careful with negative eigenvalues. Indeed -- decreasing all $|E_j^\pm|$ does not necessarily make the last sum in \eqref{5.4} negative, because $\xi^+(\beta)$ increases on $(-\infty,-1]$. This problem can be overcome if the contribution of the $\be_j^-(J)$'s to that sum is finite. Since for $\beta\uparrow -1$
\[
\xi^+(\beta)=O\bigl(|\beta+1|^3\bigr)=O\bigl(|E+2|^{\frac 32}\bigr)
\]
this means that we need
\begin{equation} \lb{5.5}
\sum_j |E_j^-+2|^{\frac 32}<\infty
\end{equation}
Then the $\liminf$ in \eqref{5.4} will be bounded from above if every change $\til J_{n-1}\to\til J_n$ decreases all $E_j^+\in (2,2+\del)$, irrespective of what happens to $E_j^-$ ($\xi^+(\beta)$ is negative on $(-\infty,-1]$). By \cite{KS},  \eqref{5.5} holds whenever $J-J_0\in L^2$.

But before we can use this idea to handle certain trace class perturbations as in Section \ref{S4}, we first need to find some $a_n,b_n$ to be perturbed. Our aim is to treat Coulomb Jacobi matrices with $2\al>\pm\be$ and show they are Szeg\H o at $\mp 2$. To prove the next result, we will return to the methods of Section \ref{S2}.

\begin{lemma} \lb{L5.3} Suppose $a_n\to 1$, $b_n\to 0$.

\begin{SL} 
\item[{\rm{(i)}}]  Let $\{a_n\}$ be eventually strictly monotone and
\begin{equation} \lb{5.6}
\frac{a_{n}-a_{n-1}}{a_{n+1}-a_n}\to 1 \qquad\qquad \frac{b_{n+1}-b_{n}}{a_{n+1}-a_n}\to \omega
\end{equation}
with $\omega$ finite. If eventually 
\[
\omega\sgn(a_{n+1}-a_n) <-2\sgn(a_{n+1}-a_n)
\]
then there are $\del>0$, $c>0$ such that $\nu'(x)\ge c\sqrt{4-x^2}$ in $(2-\del,2)$.
\item[{\rm{(ii)}}] Let $\{b_n\}$ be eventually strictly monotone and
\begin{equation} \lb{5.7}
\frac{b_{n}-b_{n-1}}{b_{n+1}-b_n}\to 1 \qquad\qquad \frac{a_{n+1}-a_{n}}{b_{n+1}-b_n}\to \omega_1
\end{equation}
with $\omega_1$ finite. If eventually 
\[
\omega_1\sgn(b_{n+1}-b_n) <-\tfrac 12\sgn(b_{n+1}-b_n)
\]
then there are $\del>0$, $c>0$ such that $\nu'(x)\ge c\sqrt{4-x^2}$ in $(2-\del,2)$.
\end{SL}
\end{lemma}

{\it Remarks.}
1. (ii) is (i) with $\omega_1=\omega^{-1}$. It handles the case $\omega=\pm\infty$.
\smallskip

2. In particular, such $J$ are Szeg\H o at 2 whenever $J-J_0\in L^2$.
\smallskip

3. By symmetry, same result holds for Szeg\H o condition at $-2$, with ``$<-2$'' and ``$<-\tfrac 12$'' replaced by ``$>2$'' and ``$<\tfrac 12$''
\smallskip

\begin{proof} (i)
First notice that \eqref{2.12} holds because $a_n$ is (eventually) monotone, and either $b_n$ is monotone (if $\omega\neq 0$) or $|b_{n+1}-b_n|\le |a_{n+1}-a_n|$ (if $|\omega|<1$). Hence, we can use Lemma \ref{L2.2}. This time we will work with $S_n$ instead of $T_n$, because it has a simpler recurrence relation \eqref{2.3}. Notice that by the proof of Lemma \ref{L2.2}, for every $|x|<2$ we have $S_n(x)\to \sqrt{4-x^2}/2\pi\nu'(x)$. The result will follow if we prove that $S_n(x)\le C$ for some $C<\infty$, all $x\in(2-\del, 2)$ and all large $n$.

We will show this by proving that for some $K$ and all large enough $n$ we have $S_{n+K-1}(x)\le S_{n-1}(x)$ for all $x\in(2-\del,2)$. That is, we will iterate \eqref{2.3} $K$ times at once. Here $K\ge 3$ and $\del$ will be fixed, but they will not be specified until later.

We let $n$ be large and such that for all $j\ge n$ we have $|a_j-1|<\del$ and $|b_j|<\del$, and we take $x\in(2-\del,2)$. Then by \eqref{1.0a} in the form \eqref{3.1} we obtain for $P_n\equiv P_n(x)$ and $k\in\{0,\dots,K-1\}$
\[
P_{n+k}=(k+1)P_n-kP_{n-1} + O(\del)(|P_n|+|P_{n-1}|)
\]
We also have 
\begin{align*}
a_{n+k+1}^2-a_{n+k}^2=& (a_{n+k+1}-a_{n+k})(2+o(1))
\\ a_{n+k}(b_{n+k+1}-b_{n+k})=& (a_{n+k+1}-a_{n+k})(\omega+o(1))
\end{align*}
with $o(1)=o(n^0)$ taken w.r.t.~$n$. From these estimates we obtain
\begin{align*}
S_{n+k}-& S_{n+k-1}= (a_{n+k+1}^2-a_{n+k}^2)P_{n+k}^2+a_{n+k}(b_{n+k+1}-b_{n+k})P_{n+k}P_{n+k-1}
\\ =& (a_{n+k+1}-a_{n+k}) \Bigl\{ \bigl[(2+o(1))(k+1)^2+(\omega+o(1))k(k+1)\bigr] P_n^2 
\\ & - \bigl[(2+o(1))2k(k+1)+(\omega+o(1))(2k^2-1)\bigr] P_nP_{n-1}
\\ & + \bigl[(2+o(1))k^2+(\omega+o(1))k(k-1)\bigr] P_{n-1}^2 + O(\del)(P_n^2+P_{n-1}^2)  \Bigr\}
\end{align*}
where the $O(\del)$ also depends on $K$ and $\omega$ (but not on $x$ or $n$). Using the identities $\sum_{k=0}^{K-1} k^2=K(2K^2-3K+1)/6$, $\sum_{k=0}^{K-1}k=K(K-1)/2$ and $a_{n+k+1}-a_{n+k}=(a_{n+1}-a_n)(1+o(1))$, we obtain for $K\ge 3$
\begin{align*}
\frac 3K \frac{S_{n+K-1}-S_{n-1}}{a_{n+1}-a_n} = & \, O(\del)(P_n^2+P_{n-1}^2)  
\\ & + \bigl[2K^2+3K+1+\omega(K^2-1)+o(1)\bigr] P_n^2 
\\ & - \bigl[4K^2-4+\omega(2K^2-3K-2)+o(1)\bigr] P_nP_{n-1} 
\\ & + \bigl[2K^2-3K+1+\omega(K^2-3K+2)+o(1)\bigr] P_{n-1}^2 
\end{align*}
where both $O(\del)$ and $o(1)$ depend on $K$ and $\omega$.
Let us denote by $I,II,III$ the three square brackets in the above expression, without the $o(1)$ terms.
If $I\cdot III-(II/2)^2>0$, then for small enough $\del$ and large $n$ (so that $O(\del)$ and $o(1)$ are negligible) the above expression will have the same sign as $I$. We have $I\cdot III-(II/2)^2>0$ whenever 
\[
\omega\notin [c_1(K),c_2(K)]\equiv \biggl[-2-\frac{6+2\sqrt{3}\sqrt{K^2-1}}{K^2-4},-2-\frac{6-2\sqrt{3}\sqrt{K^2-1}}{K^2-4} \biggr]
\]
Also, $I>0$ when $\omega>d(K)\equiv -(2K^2+3K+1)/(K^2-1)$ and $I<0$ when $\omega<d(K)$. Since $c_1(K),c_2(K),d(K)\to -2$ and by the above
\[
\sgn(S_{n+K-1}-S_{n-1})=\sgn(a_{n+1}-a_n)\sgn(I)
\]
one only needs to take $K$ large enough so that $\omega>\max\{c_2(K),d(K)\}$ (if $\sgn(a_{n+1}-a_n)<0$) or $\omega<\min\{c_1(K),d(K)\}$ (if $\sgn(a_{n+1}-a_n)>0$). Then for small enough $\del$ and all large $n$ one obtains $\sgn(S_{n+K-1}(x)-S_{n-1}(x))=-1$ whenever $x\in(2-\del, 2)$. The result follows.

(ii) The proof is as in (i), but with the role of $a_{n+1}-a_n$ played by $b_{n+1}-b_n$. We obtain $I=\omega_1(2K^2+3K+1)+K^2-1$ and $\sgn(S_{n+K-1}-S_{n-1})=\sgn(b_{n+1}-b_n)\sgn(I)$ whenever
\[
\omega_1\notin \biggl[-\frac 12-\frac{\sqrt{3}}{2\sqrt{K^2-1}}, -\frac 12+\frac{\sqrt{3}}{2\sqrt{K^2-1}}\biggr]
\]
\end{proof}

Now we are ready to introduce errors and state the main result of this section.

\begin{theorem} \lb{T5.4} Suppose $\til J$ has  
\[
\til a_n\equiv a_n + O(n^{-1-\eps}) \qquad \qquad \til b_n\equiv b_n +O(n^{-1-\eps})
\]
where $\sum_{n=1}^\infty (a_n-1)^2 + \sum_{n=1}^\infty b_n^2  <\infty$ and $\eps>0$.
\begin{SL} 
\item[{\rm{(i)}}] Assume $a_n,b_n$ satisfy \eqref{5.6} and $n^{2+\eps}|a_{n+1}-a_n|\to\infty$. If eventually
\[
\omega\sgn(a_{n+1}-a_n) < -2\sgn(a_{n+1}-a_n)
\]
then $\til J$ is Szeg\H o at $2$. If eventually
\[
\omega\sgn(a_{n+1}-a_n) > 2\sgn(a_{n+1}-a_n)
\]
then $\til J$ is Szeg\H o at $-2$.
\item[{\rm{(ii)}}] Assume $a_n,b_n$ satisfy \eqref{5.7} and $n^{2+\eps}|b_{n+1}-b_n|\to\infty$. If eventually
\[
\omega_1\sgn(b_{n+1}-b_n) < -\tfrac 12\sgn(b_{n+1}-b_n)
\]
then $\til J$ is Szeg\H o at $2$. If eventually
\[
\omega_1\sgn(b_{n+1}-b_n) < \tfrac 12\sgn(b_{n+1}-b_n)
\]
then $\til J$ is Szeg\H o at $-2$.
\end{SL}
\end{theorem}

{\it Remark.}
Notice that if $\sup \{n^{2+\eps}|a_{n+1}-a_n|\}<\infty$, then $|a_n-1|\lesssim n^{-1-\eps}$ and since (in (i)) $\omega$ is finite, we also have $|b_n|\lesssim n^{-1-\eps}$. Hence, $J-J_0$ is trace class and hence Szeg\H o by \cite{KS}.
\smallskip

\begin{proof}
(i) We follow the proof of Theorem \ref{T4.3}. First we increase $a_n$ by $6Cn^{-1-\eps}$ with $C$ from \eqref{4.0}.
We have
\begin{align*}
\frac{a_{n}+\frac{6C}{n^{1+\eps}}-a_{n-1}-\frac{6C}{(n-1)^{1+\eps}}} {a_{n+1}+\frac{6C}{(n+1)^{1+\eps}}-a_{n}-\frac{6C}{n^{1+\eps}}} 
- \frac{a_{n}-a_{n-1}}{a_{n+1}-a_n} = \frac{O(1)}{n^{2+\eps}(a_{n+1}-a_n)+O(1)}\to 0
\end{align*}
So if we call $a_n+6Cn^{-1-\eps}$ again $a_n$, we still have $(a_{n}-a_{n-1})/(a_{n+1}-a_n)\to 1$. Similarly, $(b_{n+1}-b_n)/(a_{n+1}-a_n)\to \omega$. And, of course, $\{a_n\}$ has the same type of monotonicity as before, by the assumption $n^{2+\eps}|a_{n+1}-a_n|\to\infty$. We call $J$ the matrix with these new $a_n,b_n$. By hypothesis $J-J_0\in L^2$, so $J$ is Szeg\H o at 2 by Lemma \ref{L5.3}(i) and \eqref{5.1}. 

Now we consider the same $\til J_n$ as in the proof of Theorem \ref{T4.3}. The first of them is $\til J_{N-1}$ and it is Szeg\H o at 2 because it is a finite-rank perturbation of $J$. Each next $\til J_n$ will $\del$-minorate $\til J_{n-1}$. That proves that in \eqref{5.4} (with $\til J_{N-1}$ in place of $J$) the sum involving $\beta_j^+$ will be bounded above by $K\xi^+(M)$ with $K$ and $M$ as in Lemma \ref{L4.1}. The sum with $\beta_j^-$ will be bounded above by $\sum_j(-\xi^+(\beta_j^-(\til J_{N-1})))$ and this is finite by \eqref{5.5} (which holds because $\til J_{N-1}-J_0\in L^2$). So the $\liminf$ in \eqref{5.4} cannot be $+\infty$ and the result follows.

(ii) The proof is identical.
\end{proof}

\begin{corollary} \lb{C5.5}
Let $\eps>0$ and
\begin{equation} \lb{5.8}
a_n\equiv 1+\frac \al n +O(n^{-1-\eps}) \qquad \qquad b_n\equiv \frac \be n +O(n^{-1-\eps})
\end{equation}
If $2\al>\pm \be$, then $J$ given by \eqref{1.0} is Szeg\H o at $\mp 2$.
\end{corollary}

\begin{proof}
Use Theorem \ref{T5.4}(i) (if $\al\neq 0$) or (ii) (if $\al=0$) with $a_n\equiv 1+\al/n$, $b_n\equiv \be/n$.
\end{proof}

As for other pairs $(\al,\be)$ in \eqref{5.8}, Theorem 4.4(ii) in \cite{SZ} shows that if $2\al<\pm \be$, then $J$ cannot be Szeg\H o at $\mp 2$. Hence, the $(\al,\be)$ plane is divided into 4 regions by the lines $2\al=\pm\be$. Inside the right-hand region $J$ is Szeg\H o, inside the top and bottom regions $J$ is Szeg\H o only at, respectively, $2$ and $-2$, and inside the left-hand region $J$ is Szeg\H o neither at $2$ nor at $-2$. On the borderlines the situation is as follows. If $2\al=\pm\be$ and $\al\ge 0$, then Corollary \ref{C4.4} shows that $J$ is Szeg\H o, and so Szeg\H o at both 2 and $-2$. If $2\al=\pm\be$ and $\al<0$, then $J$ cannot be Szeg\H o at $\pm 2$ by Theorem 4.4(ii) in \cite{SZ}. I think that such $J$ {\it is\/} Szeg\H o at $\mp 2$.

Finally, it should be mentioned that although we have mainly considered Coulomb behavior of $a_n,b_n$, the above picture is valid in more general setting as well. For example in the case $a_n\equiv 1+\al n^{-\gamma}+O(n^{-1-\eps})$ and $b_n\equiv \be n^{-\gamma}+O(n^{-1-\eps})$ with $\tfrac 12<\gamma\le 1$, $\eps>0$, as implied by results of \cite{SZ} and this paper.

\end{document}